\documentclass{article}

\usepackage{PRIMEarxiv}

\usepackage[utf8]{inputenc}
\usepackage[T1]{fontenc}

\usepackage{amsmath}
\usepackage{amsfonts}

\usepackage{graphicx}
\graphicspath{{media/}}

\usepackage{booktabs}
\usepackage{array}
\usepackage{longtable}
\usepackage{tabularx}

\usepackage{tikz}
\usetikzlibrary{arrows.meta}
\usetikzlibrary{positioning}

\usepackage{subcaption}

\usepackage{hyperref}
\usepackage{url}
\usepackage{microtype}
\usepackage{nicefrac}
\usepackage{todonotes}
\usepackage{ifthen}
\usepackage[misc]{ifsym}
\usepackage{lipsum}

\usepackage{listings}
\usepackage{xcolor}
\usepackage{fancyhdr}
\title{A Low-Code Approach for the Automatic Personalization of Conversational Agents
}

\author{
Aaron Conrardy \\
Luxembourg Institute of Science and Technology, Luxembourg \\
University of Luxembourg, Luxembourg \\
\texttt{aaron.conrardy@list.lu}
\and
Alfredo Capozucca \\
University of Luxembourg, Luxembourg \\
\texttt{alfredo.capozucca@uni.lu}
\and
Jordi Cabot \\
Luxembourg Institute of Science and Technology, Luxembourg \\
University of Luxembourg, Luxembourg \\
\texttt{jordi.cabot@list.lu}
}

\begin{document}

\newpage
\thispagestyle{empty}

\vspace*{\fill}

\begin{center}
\Large
\textbf{This is a preprint.}\\[1em]

The peer-reviewed paper has been published in the proceedings of ICWE 2026.\\[1em]

Please reference the published version. \\[1em]

DOI: \url{https://doi.org/10.1007/978-3-032-29372-5\_4}
\end{center}

\vspace*{\fill}

\newpage

\maketitle

\begin{abstract}
The rise of Large Language Models (LLMs) has increased the demand for Conversational Agents (CAs) capable of understanding human conversations as part of web applications. While traditional CAs consist of deterministic states, LLMs enhance their capabilities to handle open conversations, handling arbitrary requests. Numerous tools exist that allow non-technical users to create such CAs. Yet, the creation of personalized CAs able to adapt to the profile of end-users to offer an optimal user experience remains in the hands of experienced developers implementing ad-hoc personalizations. In this work, we propose a pipeline that follows a low-code/no-code approach to facilitate the modeling and generation of personalized CAs. A pilot user study was performed to get preliminary results on perceived usability and usefulness and the full pipeline has been implemented on top of an open-source low-code platform.
\end{abstract}


\bibliographystyle{plainurl}\section{Introduction}

Conversational Agents (CAs) are software applications that interact autonomously with end-users in natural language, supporting task completion, conversational web, service delivery, as well as social engagement and entertainment \cite{conversationalweb,riseofbots,slragentmetamodel}. Known under other names such as chatbots or digital assistants, they have become a ubiquitous component of modern platforms and services. While their adoption was already increasing before the widespread use of Large Language Models (LLMs), recent advances in LLMs have further established CAs as a common additional interaction modality in web applications and more. 
CAs may support different types of conversations, ranging from predefined, task-oriented dialogues based on deterministic rules to open-ended interactions enabled by LLMs, or a combination of both.

Personalized CAs (PCAs) aim to further elevate the users' experience by personalizing their interaction with the CA. Personalization refers to adapting systems to better fit an individual user’s needs, preferences, and context \cite{whatispersonalization} offering an optimal user experience.

At a high level, personalization typically consists of two processes: user profiling and system adaptation \cite{whatispersonalization} where the latter takes care of personalizing the interaction to the specific attributes of the user. For instance, for extroverted users, aligning CAs to act extroverted by adapting communication styles has shown to increase engagement and usage intention \cite{applyingpersonality}.  
While these processes are well-studied \cite{artmachines}, their implementation in CAs remains complex and requires deep technical expertise \cite{difficulttodevelop}. %

In this sense, this paper aims to facilitate the personalization of CAs, enabling non-technical people to define the type of users they want to support, the adaptations to be done for each user type and the generation of the corresponding personalized agents. To do so, we propose a new low-code pipeline to model and automatically generate PCAs. A style of model-driven development \cite{lowcodejordi,coin}, low-code development reduces the need for manual programming by relying on higher-level, graphical models to describe system behavior. 
Such approaches have already been applied to the development of CAs \cite{slragentmetamodel} and, partially, to the personalization of user interfaces \cite{mdeadaptiveaccessibility,mdeadaptiveUI} but, to the best of our knowledge, not to the challenge of creating PCAs. 

The rest of this paper is structured as follows. Section \ref{overview} presents an overview of the proposed pipeline. Section \ref{sec:modlang} presents the modeling of PCAs, followed by Section \ref{modeltrans} that describes how these PCA models are transformed into the actual agents. Tooling support is discussed in Section \ref{tool}. Section \ref{study} discusses a user study to validate our approach. Section \ref{rw} presents the related work. The paper is concluded in Section \ref{conclusion}.

\section{Overview of our low-code pipeline for the development of PCAs}\label{overview}

\begin{figure}[h]
    \centering
    \includegraphics[width=0.78\linewidth]{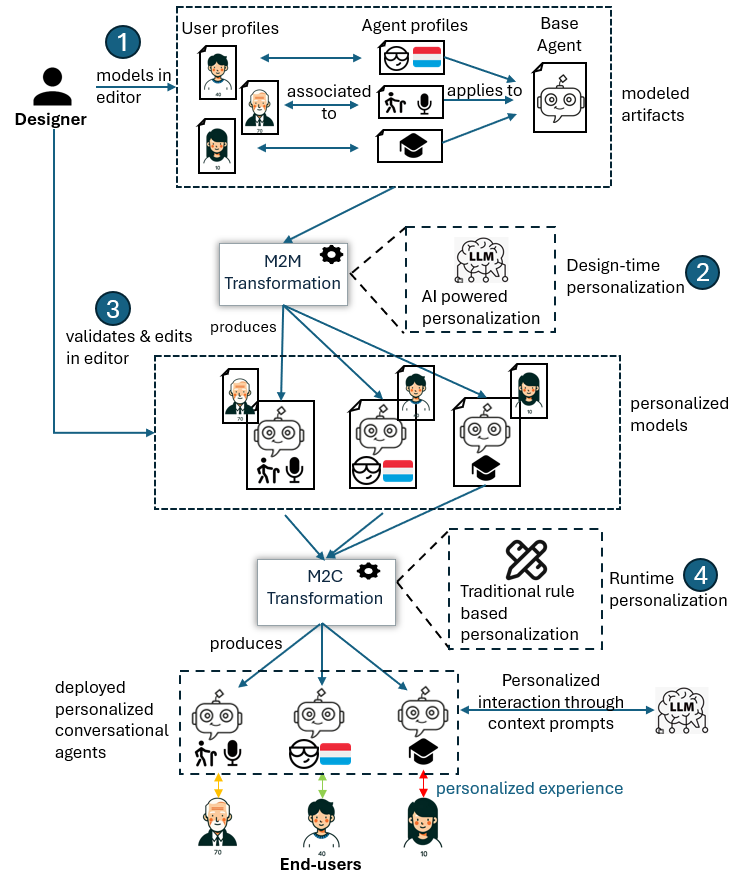}
    \caption{Overview of steps of low-code personalized agent creation pipeline}
    \label{fig:genericoverview}
\end{figure}

In this section, we present our low-code pipeline for designing and generating PCAs. Figure \ref{fig:genericoverview} illustrates the four main activities: 
\begin{enumerate}
    \item \textbf{Modeling the user and agent profiles}: The PCA designer models the user profiles (representing the types of users we target in our CA), the base agent model (with the common functionality) and the agent profiles (representing the different agent adaptations to be mapped to the corresponding user profiles). Separating the agent profile from the base agent model makes the agent profiles agent-agnostic, as they can be applied to any agent. Similarly, separating user profiles from agent profiles also enables the mapping of the same agent profile to different users, fostering reusability.  
    \item \textbf{Design time personalization}: Based on the user and agent profiles, the agent model is enriched via a model-to-model (M2M) transformation that adapts any agent answer hardcoded in the model. At this time, adaptations can include translation \cite{zhu-etal-2024-multilingual}, complexity variation \cite{complexity} and integration of cultural values \cite{culturalstuff}) among others. An LLM is used to automate these adaptations. Note that design-time adaptation is always preferred when possible as it improves runtime performance (removing the need for additional calls at every agent request) and it enables the agent validation before putting it into production.   
    \item \textbf{Manual model validation and update}: At this stage, designers can validate the personalized texts in each model and apply manual modifications if needed (e.g., improving the adapted text or adding unique states to one of the personalized agent models for a deeper adaptation). At the end of this phase, we have a set of personalized agent models linked to specific user profiles. The next step will complete the adaptation process.
    \item \textbf{Runtime personalization}: As part of a final model-to-code (M2C) transformation and, based on the specified customization in the agent profile, appropriate code excerpts are included by
    the deterministic code-generation templates. This code extends the agent behavior with additional adaptations that will be dynamically triggered when the corresponding user interacts with the agent.

\end{enumerate}

\section{Modeling users, agents, and agent profiles}\label{sec:modlang}
To support the modeling of the previously discussed artifacts, we define a set of interrelated domain-specific languages (DSLs). 
Each DSL is presented by describing both their abstract (using a metamodel expressed as a UML class diagram) and concrete syntax (using a graphical or form-based notation).

To illustrate our approach, we use as a running example a personalized gym assistant designed to provide accessible and adapted advice to different user profiles. The agent’s core task is to provide pre-defined gym-related advice on nutrition and exercises, and be able to handle any other query via generative AI as a fallback. Two user profiles are considered to receive a personalized experience: (1) an elderly user that only speaks their native language, and (2) a paraplegic user who cannot use their legs. The former prefers oral interaction, simple and formal text with a conversation in their mother tongue. The latter must not receive exercise recommendations involving the lower body, guaranteeing that adequate advice is given.

\subsection{User modeling}

User profile models have the goal to describe the end-user of an application \cite{slrusermodeling}. They can either describe individual users or user categories (e.g., elderly users in our running example) to which then real users are mapped to. 

\begin{figure}[t]
    \centering
    \includegraphics[width=0.7\linewidth]{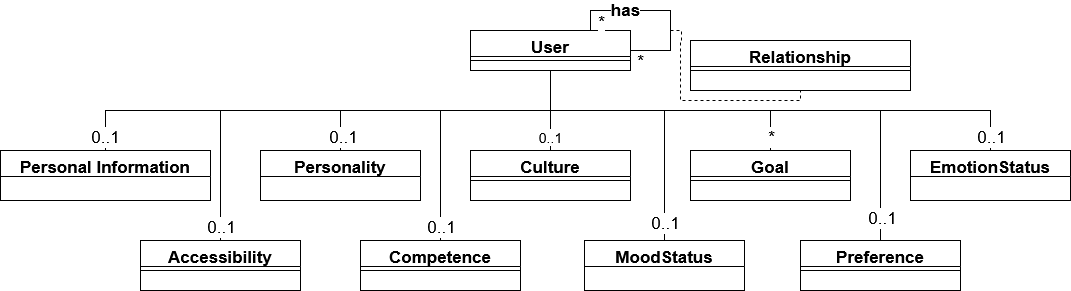}
    \caption{High-level view of user metamodel from \cite{conrardy2025unifiedusermodelinglanguage}}
    \label{fig:user_metamodel}
\end{figure}

\textbf{Abstract syntax} 
For this metamodel, we reuse the user metamodel we proposed in \cite{conrardy2025unifiedusermodelinglanguage}, with Figure \ref{fig:user_metamodel} showing a high-level view.
The metamodel consolidates results of a Systematic Literature Review (SLR) on user modeling that we performed \cite{slrusermodeling}, providing a complete and detailed overview of users via the proposed formalization, allowing the definition of user profiles with various attributes.

\textbf{Concrete syntax} Conforming to the rules of the abstract syntax, as part of this work we have developed a  graphical notation to model user profiles.

\begin{figure}
    \centering
    \includegraphics[width=0.7\linewidth]{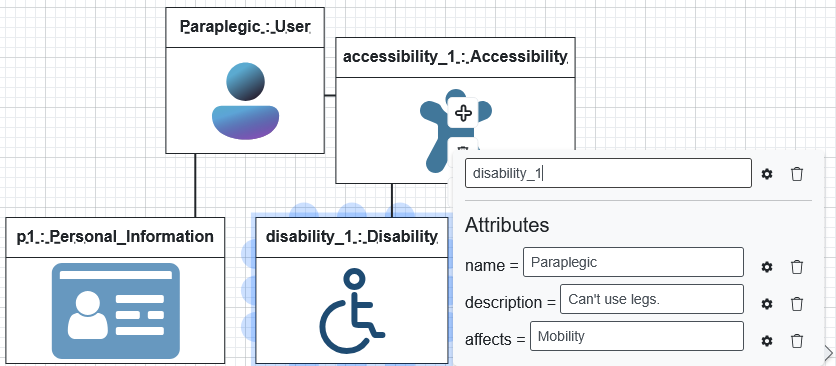}
    \caption{Paraplegic user profile model in graphical notation}
    \label{fig:paraplegic-model}
\end{figure}

Figure \ref{fig:paraplegic-model} shows a partial user profile model for our running example, representing the paraplegic user profile, containing the information on the disability and what it affects, as it is relevant for the personalization. 
The full list of icons of our graphical notation is available in our online tool (see Section \ref{tool}).

\subsection{Agent modeling}

The base agent model reflects the conversational states of the agent, thus, its core functionality. %

\begin{figure}[t]
    \centering
    \includegraphics[width=0.8\linewidth]{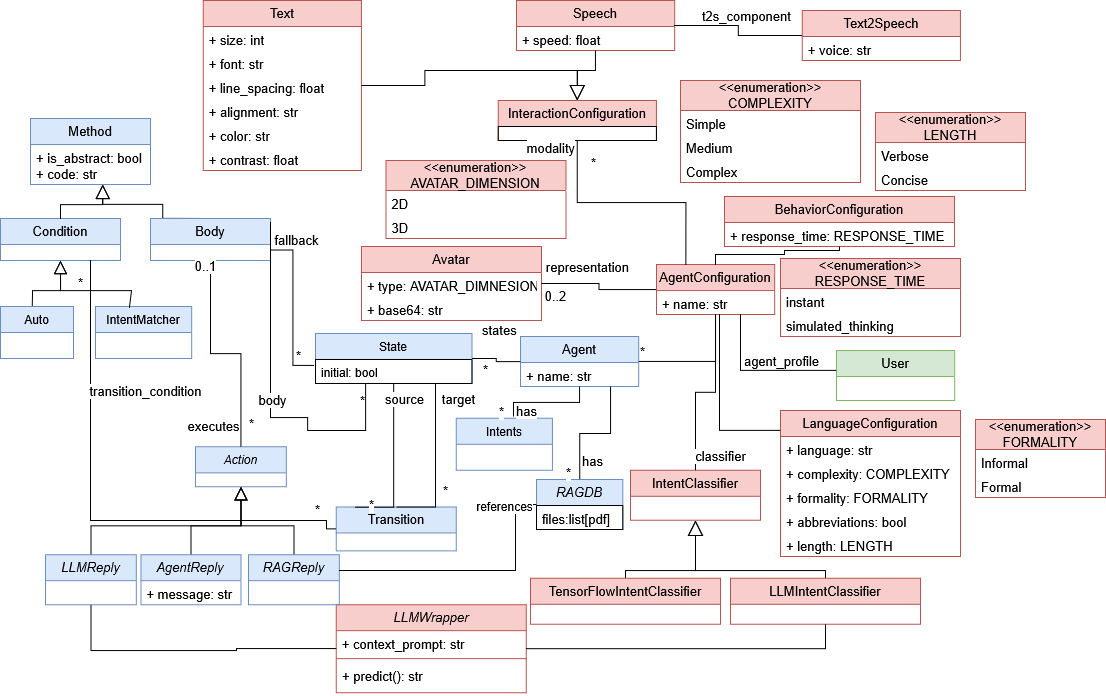}
    \caption{Agent and agent profile metamodel}
    \label{fig:agent_metamodel}
\end{figure}

\textbf{Abstract syntax} The metamodel to describe agents is shown in Figure \ref{fig:agent_metamodel} (classes in blue). Note that this is a simplified version, omitting attributes and (sub)classes not relevant to showcase the personalization pipeline, as it follows a formal-state-machine-like syntax commonly used when defining CAs \cite{fsm}.  

At its core, the metamodel includes the \textit{State}, \textit{Body} of states, executed \textit{Actions} in a given body, \textit{Transitions} that define when to jump from one state to another once specific \textit{Conditions} are fulfilled, and \textit{Intents} that represent the possible messages an end-user might pose when interacting the agent. Note that for the actions, one can choose between sending a pre-defined response, an AI generated response via an LLM, or an AI generated response using retrieval augmented generation (RAG) based on an available database.

\textbf{Concrete syntax} We propose a graphical notation that strongly follows the UML state machine formalism\footnote{https://www.omg.org/spec/UML/2.5.1/About-UML}. %

\begin{figure}[t]
    \centering
    \includegraphics[width=0.8\linewidth]{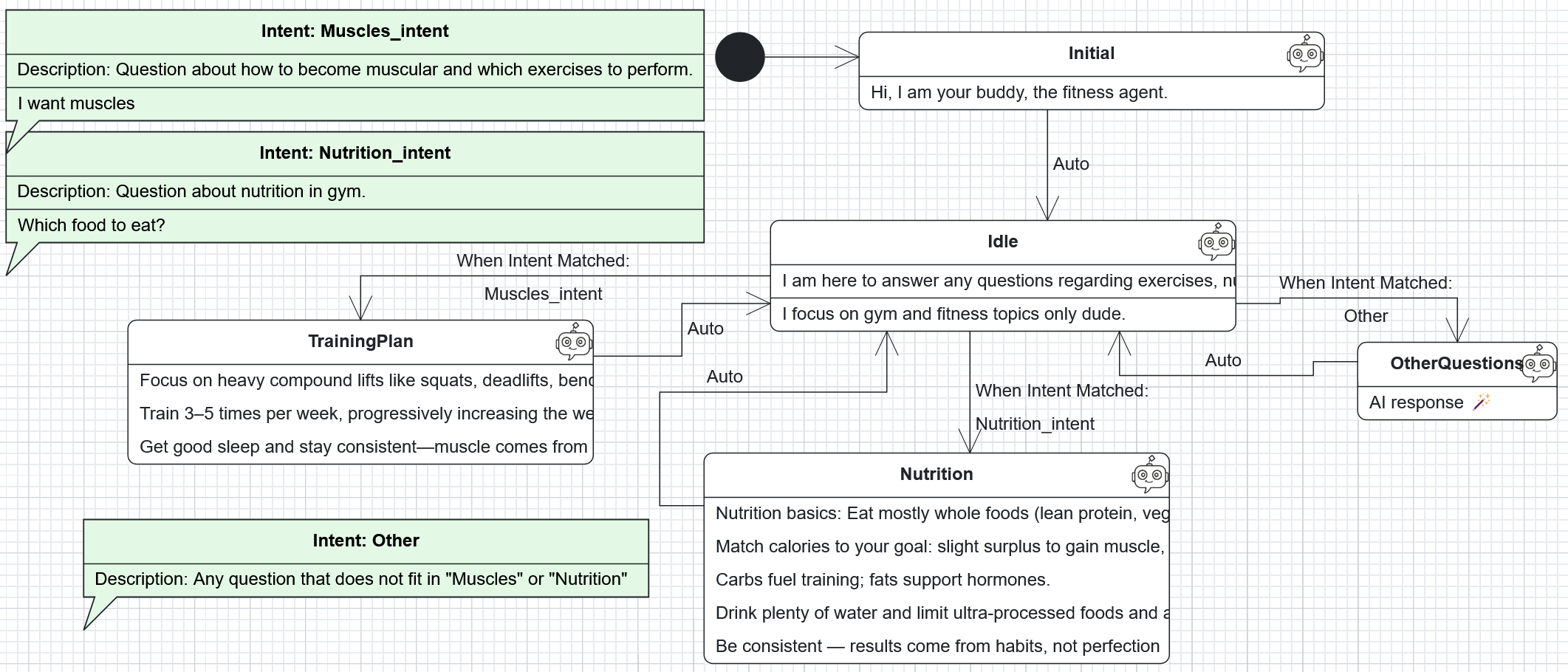}
    \caption{Graphical agent model representing the gym assistant agent}
    \label{fig:agent_diagram}
\end{figure}

Figure \ref{fig:agent_diagram} displays our running agent example using this graphical syntax. %
For simplicity, the agent answers only one question before returning to an idle state. The model consists of four states: an initial greeting state that automatically transitions to an idle state, two predefined response states (\textit{TrainingPlan} and \textit{Nutrition}), and an \textit{OtherQuestions} state handled by an LLM. Three intents (\textit{Muscles\_intent}, \textit{Nutrition\_intent}, and \textit{Other}) control transitions to the corresponding states. After each response, the agent returns to the idle state.

\subsection{Modeling agent personalizations}

The key element is to model the personalization options we want our PCA to offer to all or some of the user profiles we want to support. 

But before deciding which of the adaptations we want to include in our agent, we need to establish the set of possible adaptations available for agents and what benefits each adaptation could potentially bring. 

Unfortunately, there is not a general taxonomy of personalizable components suited for CAs. Therefore, and based on the individual personalization ideas published in the literature (e.g, see \cite{taxonomysocialcues,artmachines,powerofpersonalization,agentshealthcare}), we propose in Table \ref{tab:personalization-overview} a non-exhaustive list of personalizable components in CAs, consolidated into four distinct categories: (1) Content, (2) Presentation, (3) Behavior, and (4) Modality. These four categories are inspired by an assessment scheme that was used in an SLR on PCAs in health care \cite{agentshealthcare}, where we renamed \textit{User Interface} to \textit{Presentation}, \textit{Delivery Channel} to \textit{Modality} and \textit{Functionality} to \textit{Behavior}, and broadening their definition in what they capture. A takeaway here is that both the conversation of the agent and the UI it will take place on need to be personalized.

\begin{table*}[t]
\centering
\small
\begin{tabularx}{\textwidth}{
>{\raggedright\arraybackslash\hsize=1.1\hsize}X
>{\raggedright\arraybackslash\hsize=0.6\hsize}X
>{\raggedright\arraybackslash\hsize=1.3\hsize}X}
\toprule
\textbf{Category (definition)} & \textbf{Aspect} & \textbf{Example (with reference)} \\
\midrule

\textbf{Content} 
(adapts \emph{what} is shown or done, i.e., the semantics of the interaction) 
& Selection / structure 
& Adaptive conversation paths \cite{agentshealthcare}; paragraph length adapted to cognitive capabilities \cite{artmachines} \\

& Enrichment 
& Response tailored to fit user constraints  \cite{agentshealthcare} \\

\midrule

\textbf{Presentation} 
(adapts \emph{how} content is delivered without changing its meaning) 
& Language and style 
& Code-switching \cite{artmachines}; simplified text \cite{taxonomysocialcues}; text formality \cite{powerofpersonalization,artmachines} \\

& Visual / auditory form 
& Larger text for elderly users \cite{mdeadaptiveaccessibility}; emotional tone of voice \cite{artmachines}; avatar appearance \cite{taxonomysocialcues} \\

\midrule

\textbf{Behavior} 
(adapts \emph{how the interaction unfolds} over time) 
& Interaction management 
& Response timing, turn-taking, and proactiveness \cite{taxonomysocialcues,artmachines} \\

& Social and repair strategies 
& Clarification questions \cite{artmachines}; small talk or humor \cite{taxonomysocialcues} \\

\midrule

\textbf{Modality} 
(adapts the \emph{input/output channels} used for interaction) 
& Channels 
& Text, speech, images/video, gestures, or haptics \cite{artmachines,agentshealthcare} \\

\bottomrule
\end{tabularx}

\caption{Overview of personalization dimensions with embedded definitions and representative examples.}
\label{tab:personalization-overview}
\end{table*}

\textbf{Abstract syntax} Based on the taxonomy in Table \ref{tab:personalization-overview}, we have defined the metamodel for modeling agent configurations, visible in Figure \ref{fig:agent_metamodel} (red classes).
Related to Modality, \textit{InteractionConfiguration} defines the used modalities. For Presentation, \textit{Text} and \textit{Speech} allow for the specification of stylistic changes related to the chosen channel of communication. \textit{Avatar} defines the visual representation of the CA. \textit{LanguageConfiguration} relates to stylistic changes in the used language. Related to Behavior, \textit{BehaviorConfiguration} lets users change if the agent pretends to type out a message or instantly responds. For Content, it is enough that the information of the user profile is provided, as the semantics will be changed based on that information, which is represented by the association between the agent configuration and the user. 
Additionally, the components \textit{IntentClassifier} (necessary for natural language understanding), \textit{LLMWrapper}, \textit{RagDB}, \textit{Platform} and \textit{Text2Speech} define technological configurations of the agent, focusing on used tools and methods to run the agent.

\begin{figure}[t]
    \centering
    \includegraphics[width=0.9\linewidth]{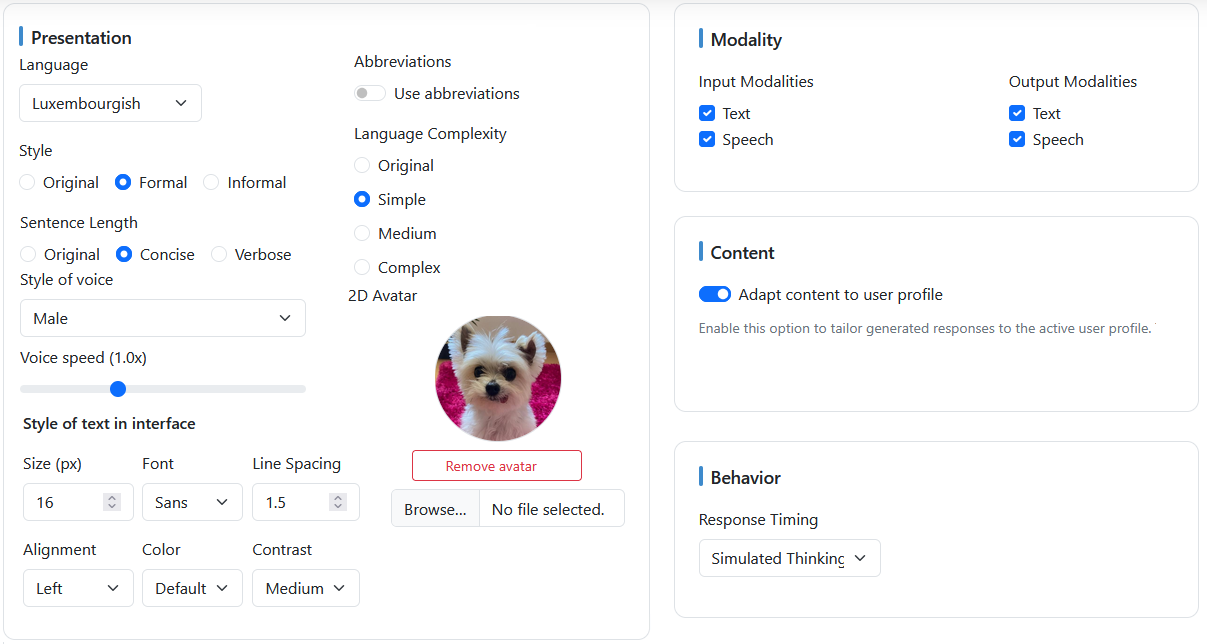}
    \caption{Form-based configuration specification}
    \label{fig:agent_configuration_edlerly}
\end{figure}

\textbf{Concrete syntax} To specify the agent profile, we decided to implement it via a form-based syntax, given the configurational and feature-like nature of the agent configuration.
Figure \ref{fig:agent_configuration_edlerly} contains the agent configuration in the form-based syntax created based on the requirements for the Elderly profile, where the four personalization categories are represented. The configurable content in each pillar is straightforward, such as setting Language Complexity to \textit{Simple} in Presentation, selecting \textit{Speech} as modalities for both input and output in Modality, or setting the boolean to adapt content to the user profile to true.

\subsection{Linking the models}
In a final step, the personalization model needs to be mapped to the affected base agent model and target user profile, illustrated in Figure \ref{fig:agent_metamodel} in the metamodel with the association starting from the class \textit{Agent}, respectively \textit{User} to \textit{Agent Configuration}. This straightforward connection is implemented via forms, where the designer defines the mapping that is then stored on top of the defined models.

\section{Generating the personalized conversational agents}\label{modeltrans}
We will now explain the two-step personalization process of our pipeline, depending on whether the personalization can be completely done at design time or it is part of the runtime execution of the agent.

This process is repeated for every agent profile we want to add to our base agent model, resulting in a family of related agents that share the same core behavior but present it on different shapes and styles depending on the personalization applied to generate each of them. 

\subsection{Design time personalization}
Design time personalization in the model-to-model transformation is only applied to the predefined textual responses in the agent model, and only if the agent profile specifies changes related to nondeterministic personalization aspects. These include changes in \textit{Language}, \textit{Style}, \textit{Sentence Length}, \textit{Abbreviations}, \textit{Language Complexity} and adapting the \textit{Content} to fit the user profile. For each aspect, an external LLM is prompted to adapt all the predefined messages to fit the specific requested adaptation for the task. 

Personalization aspects are applied sequentially: after each step, the resulting output replaces the previous version and is used as input to the next aspect. This sequential design constrains each LLM call to a single aspect, reducing prompt complexity and limiting cross‑interference between aspects.
Listing \ref{prompt} reflects the used prompt when \textit{Sentence Length} is set to concise, with other prompts following a similar structure, available in the replication package\footnote{\url{https://github.com/Aran30/LowCodeConversationalAgentPersonalization_replication}}. 
This process outputs the base agent model with the personalized textual responses for a given agent configuration.

\subsection{Runtime personalization}
Runtime personalization requires generating code that guarantees a personalized runtime interaction. For each personalization aspect, deterministic rules are applied, adding the necessary bits of code (de)activating or modifying certain agent features or wrapping agent calls to external services to make sure the call response is adapted to the personalization requirement. 

Among the former we have \textit{Style of Voice}, \textit{Voice Speed}, \textit{Style of text in interface}, \textit{Avatar}, \textit{Input and output Modality} and \textit{Response Timing}.
For the latter, a context prompt will be added describing the expected personalization aspects the LLM should adhere to when preparing the response. If necessary, a call to a second LLM to either confirm the adaptation is correct (e.g. checking the content is suitable for the target age) or forcing a reiteration of the adaptation (e.g. if the second LLM detects the adaptation is insufficient) can be also added. 

\begin{figure}[t]
    \centering
    \includegraphics[width=\textwidth]{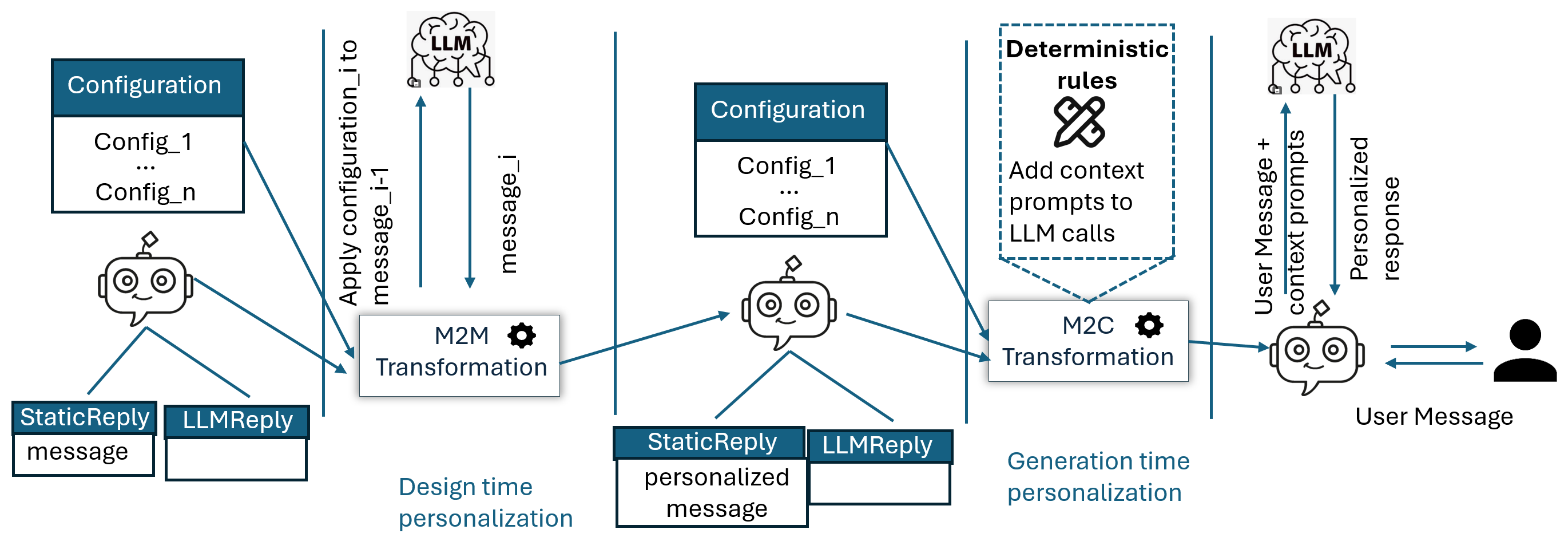}
    \caption{Process for nondeterministic personalization aspects}
    \label{fig:message-personalization-process}
\end{figure}

Figure \ref{fig:message-personalization-process} illustrates some personalization aspects for our running example.

\begin{figure}
\begin{lstlisting}[frame=single,breaklines=true,basicstyle=\small\ttfamily]
You are an editing engine focused on brevity. Rewrite each numbered text to be concise while preserving the original meaning. Remove redundancy, trim filler, and keep sentences short. Return the rewritten texts as a numbered list in the same order.
\end{lstlisting}
\caption{Prompt for concise sentence generation.}
\label{prompt}
\end{figure}

\section{Tool support}\label{tool}
The complete pipeline has been implemented by extending
the open source BESSER framework \cite{alfonso2024building}, which focuses on the low-code/no-code development of smart software. 
Note that the screenshots shown in Section \ref{sec:modlang} were taken directly from the tool. 
More precisely, we extended three components of the BESSER suite\footnote{\url{https://github.com/BESSER-PEARL}}, which are all grouped in the replication repository\footnote{\url{https://github.com/Aran30/LowCodeConversationalAgentPersonalization_replication}}.

\begin{figure}[t]
    \centering
        \includegraphics[width=0.9\linewidth]{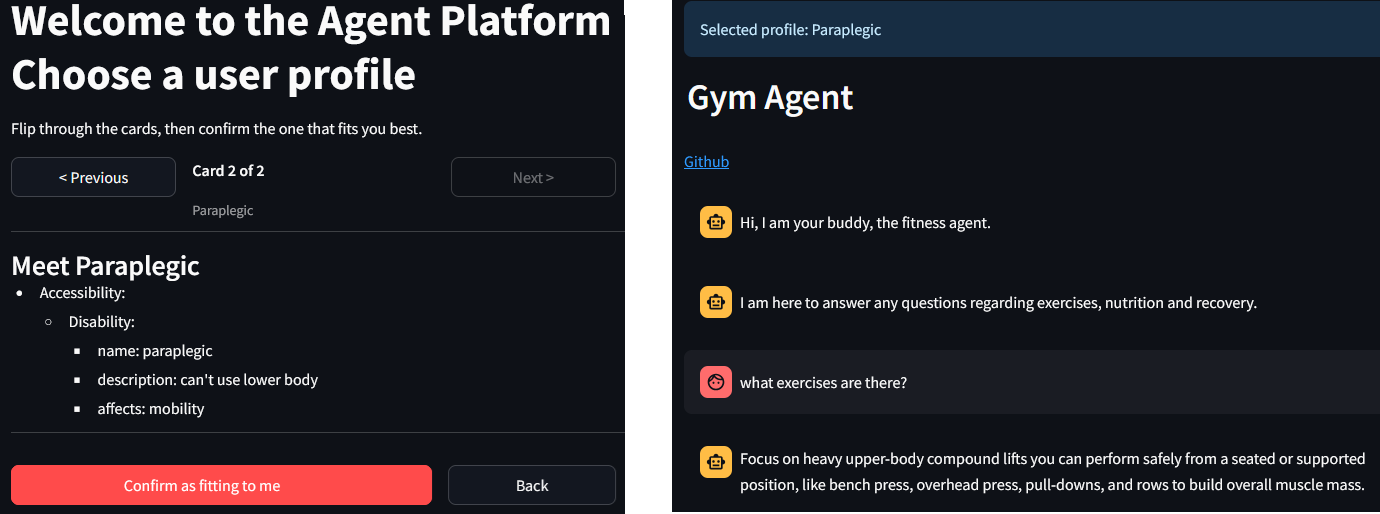}
        \caption{Profile picker and chat with personalized agent for paraplegic user}
        \label{fig:paraplegicchat}

\end{figure}

First, we extended the BESSER Agentic Framework (BAF). In itself, the framework acts as a Python-based library to create CAs following a state-machine-like formalism. We extended it to be able to, at design time, specify agent variations of state actions and user profile, and at runtime, allow end-users to choose a profile to trigger the corresponding personalized interactions, visible in Figure \ref{fig:paraplegicchat}. 

Second, we extended the core low-code BESSER platform. In particular, we added new transformations and generators to implement the personalizations as described in the Section \ref{modeltrans}. %

Finally, we added new modeling perspectives to the BESSER Web Modeling editor to enable the graphical modeling of the profiles and the form-based configuration for the personalization.

\section{Pilot Study}\label{study}

A pilot user study was conducted to evaluate our CA personalization tool regarding the following two research questions (RQs): perceived usability (RQ1) and perceived usefulness (RQ2). The study further aimed to explore whether users with different technical backgrounds could understand and complete the personalization workflow. The previously shared replication package also contains the study handout and questionnaire data. 

\textbf{Participants} Eight participants were recruited for the exploratory study through direct contact to obtain the required profiles. Four participants had a technical background with prior experience in software development and modeling, while the other four had no programming experience (with the exception of one who reported limited experience) and no prior exposure to software modeling. This diversity was intentional, as the study aimed to explore early perceptions across the two target user profiles of the tool: expert (technical) and layman (non-technical). 

\textbf{Procedure} After signing a consent form, participants received a document describing the running example and the tasks required to complete one iteration of the CA personalization pipeline. Upon opening the web-based modeling tool, they were presented with a pre-modeled gym agent, which they inspected and explained to the study supervisor to confirm their understanding of the model’s syntax and semantics.
Next, based on task descriptions, participants created two predefined user profiles (\textit{Elderly} and \textit{Paraplegic}), defined two agent profiles, mapped user profile to agent profile, and triggered the M2M transformation to generate personalized models. Participants also compared the adapted and original models to identify the differences that occurred through the design time adaptations.
Finally, they generated and interacted with the personalized agent. The tasks were designed to capture participants’ initial understanding and impressions of the personalization process. %
Afterward, participants completed a questionnaire including the System Usability Scale (SUS) \cite{brooke1996sus} and perceived usefulness and ease-of-use items adapted from the Technology Acceptance Model (TAM) \cite{davis1989tam}, both widely used and validated instruments for measuring usability and technology acceptance. Responses were collected on a five-point Likert scale ranging from 1 ("Strongly Disagree") to 5 ("Strongly Agree").

\textbf{Results} Table \ref{tab:evaluation} contains the result of the questionnaire, the completion time, and the intervention count as averages based on the participants' profile. The answers from the questionnaires are normalized to 100 to match the standardized SUS score.

\begin{table}[t]
\caption{Comparison of expert and layman evaluation results}
\label{tab:evaluation}
\centering
\begin{tabular}{lcc}
\hline
\textbf{Metric} & \textbf{Experts} & \textbf{Laymen} \\
\hline
SUS Score & 92.50 & 78.75\\
Perceived Usefulness (PU) & 95.00 & 85.00 \\
Perceived Ease of Use (PEOU) & 96.60 & 71.60 \\
\hline
Average completion time (min) & 39.75 & 43.25 \\
Average number of interventions (number) & 2.25 & 3.75 \\
\hline
\end{tabular}
\end{table}

\textbf{RQ1 (Usability)} The high average scores that reflect usability (SUS and PEOU) for both experts (SUS: 92.50, PEOU: 96.60) and laymen (SUS: 78.75, PEOU: 71.60) indicate good usability overall. Nonetheless, a clear difference in perception between experts and laymen ($\Delta$SUS: 13.75, $\Delta$PEOU: 25.00) is observed. A possible explanation would be the lack of familiarity with modeling tools for the laymen compared to the experts. 
In contrast, the delta between experts and laymen for completion time ($\Delta$3.50) or interventions ($\Delta$1.50) is very low, showcasing a similar performance. 
Independent of the participant profile, minor usability concerns have been noted. Most notably, we observed that it was unclear sometimes that elements in the user profile editor needed to be linked with each other, and the agent configuration page was perceived as overloaded, which increased the likelihood of overlooking elements. Addressing these and adding better documentation would increase usability.

\textbf{RQ2 (Usefulness)} Participants of both profiles perceived the pipeline to have a high usefulness, as evident by the high scores in the questionnaire (PU: 95.00 for experts and 85.00 for laymen). A small difference is observable between the profiles ($\Delta$PU: 10), which possibly stems from modeling experts understanding and appreciating the advantages of the approach more than laymen.

\textbf{Threats to validity} Several factors may threaten the validity of the claims made in this study. Regarding internal validity, as participants were acquaintances of the study supervisor, there is a risk of response bias towards more positive evaluations, although they were explicitly encouraged to provide honest and critical feedback. Concerning construct validity, the instruction sheet may have provided substantial guidance, potentially reducing the perceived complexity. Yet, we argue this reflects the documentation typically provided in production-ready tools. For external and conclusion validity, the small sample size limits the generalizability of the results and the statistical power, meaning observed trends should be interpreted with caution and confirmed in larger replications.

\section{Related Work}\label{rw}
We comment on two types of related works: 
those focusing on a model-driven approach for the generation of CA and those proposing ad-hoc adaptation techniques for CAs. As we discuss, there are no model-driven approaches that target the automatic personalization of CAs that could be embedded in a web application. 

\textbf{Low-code and model-driven approaches for developing CAs} 
Model-driven approaches have been applied in the past to the development of web and mobile UIs, see for instance, tools and languages such as IFML \cite{ifml}, OOWS \cite{Fons2008} or UWE \cite{Koch2008} among many others. Yet, these do not include a CA perspective, their support for specifying user profiles is limited to key-value pairs and the offered personalization is limited. Newer approaches focus more on personalization \cite{mdeadaptiveUI,mdeadaptiveaccessibility}, albeit not targeting either CAs and only providing textual syntaxes, therefore not catered towards non-technical users.

There are also a few approaches targeting low-code development for CAs (see this SLR conducted by Ouaddi et al. \cite{slragentmetamodel}), though, similar to the approaches above, they do not cover personalization aspects. As an example, a recent no-code approach was proposed to enable the creation of voice CAs on top of existing web-pages as a plug-in \cite{voicebot}. The approach does not focus on personalization or catering to diverse user profiles.

\textbf{Personalization in agents} Personalizing CAs is not a novel topic in itself, as evident by the existence of SLRs covering this topic (e.g., \cite{powerofpersonalization,agentshealthcare} to name a few) that provide insights about techniques that exist or types of personalization and their benefits.
Yet, tools that facilitate the creation of such personalized agents combining various existing personalization options are lacking. Our pipeline acts as an initial attempt to facilitate the creation of personalized CAs by leveraging technologies that showed positive results in the past in a way that non-technical users could benefit from.

\section{Conclusions}\label{conclusion}

This paper presented a pipeline for designing and generating personalized conversational agents using low-code and no-code principles. %
Our proposal enables the specification of user-specific agent profiles, linking users with customized agents derived from a core agent specification via a combination of deterministic transformations and LLM customizations.

As further work, we plan to increase the types of personalizations available and automatically propose personalization templates for the most common types of users. We also plan to extend the personalization to other types of UIs (e.g., for extended reality scenarios) and add some quality checks to the configuration process to detect potentially wrong or inconsistent personalization options. Finally, larger scale user studies shall be conducted to further validate the pipeline's usability, but also scalability and quality of generated personalization.

\section*{Acknowledgments}
This work is supported by the Luxembourg National Research Fund (FNR) PEARL program, grant agreement 16544475.


\bibliography{bib}

\end{document}